\documentclass[letterpaper]{article} 
\usepackage{aaai2026}  
\usepackage{times}  
\usepackage{helvet}  
\usepackage{courier}  
\usepackage[hyphens]{url}  
\usepackage{graphicx} 
\usepackage{natbib}  
\usepackage{caption} 
\usepackage{xr}
\usepackage{algorithm}
\usepackage[noend]{algpseudocode}
\usepackage{amsmath}
\usepackage[table]{xcolor}
\usepackage{multirow}
\usepackage{booktabs}
\usepackage{tabularx}
\usepackage{makecell}
\usepackage{subcaption}
\definecolor{purple}{RGB}{153,1,254}
\definecolor{orange}{RGB}{234,79,1}
\definecolor{myorange1}{HTML}{FFF5E0} 
\definecolor{myorange2}{HTML}{FFE0B2} 
\definecolor{myorange3}{HTML}{FFCC80} 
\definecolor{myorange4}{HTML}{FFB74D} 
\definecolor{myorange5}{HTML}{FFA726} 

\usepackage{newfloat}
\usepackage{listings}
\DeclareCaptionStyle{ruled}{labelfont=normalfont,labelsep=colon,strut=off} 
\floatstyle{ruled}
\newfloat{listing}{tb}{lst}{}
\floatname{listing}{Listing}
\title{BiCA: Effective Biomedical Dense Retrieval with Citation-Aware Hard Negatives}
\author{
    Aarush Sinha\textsuperscript{\rm 1}\thanks{Worked done as a UG student and currently at University of Copenhagen, Denmark}\equalcontrib,
    Pavan Kumar S\textsuperscript{\rm 2,3}\equalcontrib,
    Roshan Balaji\textsuperscript{\rm 2,3},
    Nirav Pravinbhai Bhatt\textsuperscript{\rm 2,3}\thanks{Corresponding Author}
}

\affiliations{
    \textsuperscript{\rm 1}Vellore Institute of Technology (VIT) Chennai, India\\
    \textsuperscript{\rm 2}BioSystems Engineering and Control (BiSECt) Lab, Department of Biotechnology and \\Wadhwani School of Data Science and AI, Indian Institute of Technology (IIT) Madras, Tamil Nadu India\\
    \textsuperscript{\rm 3}The Centre for Integrative Biology and Systems medicinE (IBSE), IIT Madras, Chennai, Tamil Nadu, India\\
   aarush.sinha@gmail.com, \{bt19d200,bt21d401,niravbhatt\}@smail.iitm.ac.in
}

\usepackage{bibentry}

\begin{document}

\maketitle

\begin{abstract}
Hard negatives are essential for training effective retrieval models. Hard-negative mining typically relies on ranking documents using cross-encoders or static embedding models based on similarity metrics such as cosine distance. Hard negative mining becomes challenging for biomedical and scientific domains due to the difficulty in distinguishing between source and hard negative documents. However, referenced documents naturally share contextual relevance with the source document but are not duplicates, making them well-suited as hard negatives.
In this work, we propose BiCA:  Biomedical Dense Retrieval with Citation-Aware Hard Negatives, an approach for hard-negative mining by utilizing citation links in 20,000 PubMed articles for improving a domain-specific small dense retriever.  We fine-tune the GTE$_\text{small}$ and GTE$_\text{Base}$ models using these citation-informed negatives and observe consistent improvements in zero-shot dense retrieval using nDCG@10 for both in-domain and out-of-domain tasks on BEIR and outperform baselines on long-tailed topics in LoTTE using Success@5. Our findings highlight the potential of leveraging document link structure to generate highly informative negatives, enabling state-of-the-art performance with minimal fine-tuning and demonstrating a path towards highly data-efficient domain adaptation.
\end{abstract}

\begin{links}
    \link{Code}{github.com/bisect-group/BiCA}
    \link{Datasets}{huggingface.co/collections/bisectgroup/bica-aaai26}
\end{links}

\begin{figure*}[!ht]
    \centering
    \includegraphics[width=\linewidth]{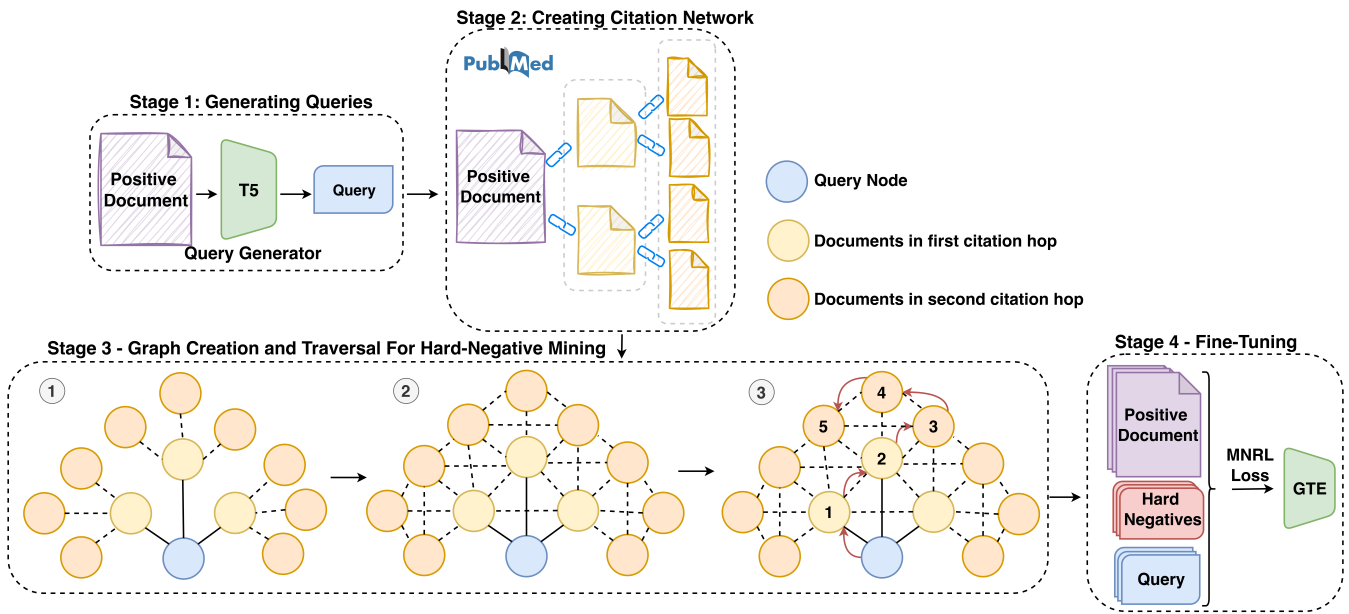}
    \caption{Our four-stage data generation and training pipeline. 
        \textbf{Stage 1:} A query is synthetically generated from a positive document's abstract using a T5 model. 
        \textbf{Stage 2:} A 2-hop citation neighborhood is constructed by retrieving papers cited by the positive document (1-hop) and papers cited by them (2-hop) via the PubMed API. 
        \textbf{Stage 3:} Hard negatives are mined via semantic graph traversal. First, similarities are computed between the query and 1-hop documents. Second, a dense, pairwise similarity graph is built for all 1-hop and 2-hop documents. Third, a 5-step greedy traversal is initiated from the 1-hop document most similar to the query, creating a path of five hard negatives. 
        \textbf{Stage 4:} The resulting (Query, Positive Document, Hard Negatives) triplet is used to fine-tune the GTE model using the multiple negative ranking loss.}
        \label{fig:cite}
\end{figure*}

\section{Introduction}

Information Retrieval (IR) is a fundamental discipline focused on extracting relevant information from vast collections of unstructured data, primarily text. IR systems employ various algorithms to match user queries with pertinent documents, integrating both exact lexical matching and semantic understanding techniques. These systems are essential for search engines, digital libraries, and question-answering applications, enabling users to efficiently navigate large volumes of information~\cite{manning2009introduction}.

Despite these advances, retrieving precise information from the rapidly expanding biomedical literature indexed in PubMed~\cite{sayers_dataBase_2011} remains a significant challenge. This difficulty is often compounded by the prevalence of low-quality, keyword based queries which may lack the specificity required to pinpoint relevant documents within such a specialized and nuanced domain. To address this issue, we propose an effective alternative by taking advantage of advanced training strategies and model architectures tailored for this complex environment.

One such strategy is Hard Negative mining, which involves selecting challenging examples that closely resemble positive samples yet are ultimately irrelevant~\cite{hnchallenge,yang2024trisamplerbetternegativesampling}. By compelling models to learn finer-grained distinctions between these difficult-to-distinguish negatives and true positives, the resulting systems exhibit more accurate rankings and improved retrieval effectiveness. Specifically for biomedical IR, the challenge lies not only in the sheer volume of literature but also in the intricate terminology and the subtle semantic relationships between concepts.

In this work, we introduce BiCA (Biomedical Citation-Aware) retrievers, a family of models designed to enhance biomedical information retrieval and out-of-domain retrieval. We propose a novel hard negative mining technique based on multi-hop citation chains within the PubMed database. This approach, combined with efficient model architectures, allows us to develop systems that are not only highly effective but also computationally efficient. We demonstrate that our models, BiCA$_\text{Base}$ and the significantly smaller BiCA$_\text{small}$, achieve state-of-the-art or highly competitive results on several biomedical and general-domain IR benchmarks, often outperforming models that are substantially larger.

\subsection{Our Contributions}
The main contributions of this work are as follows:
\begin{itemize}
 \item We introduce a novel hard negative mining strategy that constructs multi-hop citation chains from PubMed, using the \texttt{pubmed-parser}, to generate high-quality, challenging negative examples for training  retrieval models for biomedical domains.
    \item We introduce BiCA$_\text{Base}$ (110M parameters) and BiCA$_\text{small}$ (33M parameters), two dense retrieval models specifically tailored for the biomedical domain using the proposed citation-aware hard negative mining, which also demonstrate strong performance on general domain retrieval tasks.
   
    \item Extensive zero-shot evaluations of our BiCA models on 14 BEIR tasks and 4 LoTTE tasks, outperforming all baselines on several tasks in BEIR and all sub-topics on LoTTE.
    \item We provide a detailed latency analysis, demonstrating the practical efficiency of our models, particularly BiCA$_\text{small}$, on a single V100 GPU, highlighting their suitability for real-world deployment.
\end{itemize}

\section{Related Work}

\subsection{Biomedical Information Retrieval}

Recent advancements in biomedical IR have focused on integrating novel methods and leveraging large-scale data to enhance retrieval performance. One such approach is Bibliometric Data Fusion \cite{10265867}, which incorporates bibliometric metadata such as citation counts and altmetrics into retrieval systems. By using these implicit relevance signals, this method aims to improve retrieval performance, particularly for patient users, without relying on explicit relevance labeling.

A more recent development, Self-Learning Hypothetical Document Embeddings (SL-HyDE) \cite{li2024automireffectivezeroshotmedical}, introduces a zero-shot approach to medical IR by utilizing large language models (LLMs) to generate hypothetical documents based on a given query. This framework, which self-learns both pseudo-document generation and retrieval processes, improves retrieval accuracy without needing labeled data. The approach has shown notable performance across various LLM and retriever configurations, indicating its potential for enhancing zero-shot retrieval tasks.

Another important contribution to biomedical IR is the development of Neural Retrievers (NRs) \cite{Luo_Mitra_Gokhale_Baral_2022}, which address data scarcity in the biomedical domain. By proposing a template based question generation method and introducing pre-training tasks aligned with the downstream retrieval task, NRs have made substantial strides. The ``Poly-DPR" model, which encodes each context into multiple vectors, has been particularly effective, outperforming traditional methods like BM25 in certain retrieval settings.

 Finally, MedCPT \cite{jin_medcpt_2023} employs contrastive pre-training to enhance zero-shot retrieval for biomedical information. Leveraging a large collection of user click logs from PubMed, MedCPT utilizes contrastive learning to train an integrated retriever and re-ranker model. This methodology has set new state-of-the-art benchmarks, outperforming several Baselines, including larger models like GPT-3-sized cpt-text-XL.

\subsection{Biomedical Language Models}

The development of domain-specific language models has addressed the unique challenges posed by biomedical text. Models like SciFive \cite{phan_scifive_2021}, BioMegatron \cite{shin_biomegatron_2020}, and PubMedBERT \cite{gu_domain-specific_2021} have been trained on extensive biomedical corpora, enabling them to better understand specialized language and concepts. Additionally, other models such as BioBERT \cite{10.1093/bioinformatics/btz682}, PMC-LLaMA \cite{10.1093/jamia/ocae045}, ELECTRAMed \cite{miolo2021electramednewpretrainedlanguage}, BioBART \cite{yuan-etal-2022-biobart}, and BioMedLM \cite{bolton2024biomedlm27bparameterlanguage} have significantly advanced biomedical text mining and natural language processing (NLP). 

Recent advancements in biomedical language modeling have explored graph-based approaches to represent biomedical literature as knowledge graphs, effectively capturing complex relationships among entities and concepts. These knowledge graphs enhance accuracy by providing a structured framework that reflects the intricate interconnections inherent in biomedical data. Works of \cite{saxena_improving_2020},\cite{yasunaga_linkBERT_2022} and \cite{yasunaga2022dragon} show significant improvements in question-answering systems and biomedical text understanding using knowledge graphs and multi-hop frameworks.

\begin{table*}[!ht]
\centering
\small 
\setlength{\tabcolsep}{2pt} 
\begin{tabular}{@{}lccccccccccccccc|cc@{}} 
\toprule
\rotatebox{90}{\textbf{Model}} &
\rotatebox{90}{\textbf{Size}} &
\rotatebox{90}{\textbf{COVID}} &
\rotatebox{90}{\textbf{NFC}} &
\rotatebox{90}{\textbf{SCIFACT}} &
\rotatebox{90}{\textbf{SCIDOCS}} &
\rotatebox{90}{\textbf{QUORA}} &
\rotatebox{90}{\textbf{ArguAna}} &
\rotatebox{90}{\textbf{Climate-Fever}} &
\rotatebox{90}{\textbf{NQ}} &
\rotatebox{90}{\textbf{CQADup}} &
\rotatebox{90}{\textbf{DBPedia}} &
\rotatebox{90}{\textbf{Touché-2020}} &
\rotatebox{90}{\textbf{HotpotQA}} &
\rotatebox{90}{\textbf{FEVER}} &
\rotatebox{90}{\textbf{FiQA}} &
\rotatebox{90}{\textbf{Avg.}} & 
\rotatebox{90}{\textbf{Macro Avg.}} \\
\midrule
TAS-B & 66M & 0.481 & 0.319 & 0.643 & 0.149 & 0.835 & 0.434 & 0.221 & 0.463 & 0.315 & 0.384 & 0.162 & 0.584 & 0.700 & 0.300 & 0.399 & 0.399 \\ 
R-GPL & 66M & 0.760 & 0.342 & 0.678 & 0.162 & 0.808 & 0.464 & 0.231 & 0.504 & 0.348 & 0.381 & 0.264 & 0.567 & 0.791 & 0.336 & 0.474 & 0.474 \\ 
GPL & 66*5M & 0.700 & 0.345 & 0.674 & 0.169 & 0.832 & 0.483 & 0.227 & 0.467 & 0.345 & 0.360 & 0.266 & 0.636 & 0.758 & 0.344 & 0.472 & 0.472 \\ 
DPR & 110M & 0.332 & 0.189 & 0.318 & 0.077 & 0.248 & 0.175 & 0.148 & 0.474 & 0.153 & 0.263 & 0.131 & 0.456 & 0.562 & 0.112 & 0.274 & 0.274 \\ 
ANCE & 110M & 0.650 & 0.230 & 0.507 & 0.122 & 0.852 & 0.415 & 0.198 & 0.446 & 0.296 & 0.281 & 0.240 & 0.584 & 0.669 & 0.295 & 0.414 & 0.414 \\ 
Contriever & 110M & 0.596 & 0.328 & 0.677 & 0.165 & 0.865 & 0.446 & 0.237 & 0.495 & 0.284 & 0.413 & 0.230 & 0.638 & 0.758 & 0.329 & 0.463 & 0.463 \\ 
ColBERT & 110M & 0.677 & 0.305 & 0.671 & 0.145 & 0.854 & 0.233 & 0.184 & 0.524 & 0.350 & 0.392 & 0.202 & 0.593 & 0.771 & 0.317 & 0.445 & 0.445 \\ 
ColBERTv2 & 110M & 0.738 & 0.338 & 0.693 & 0.154 & 0.852 & 0.463 & 0.176 & 0.562 & 0.359 & \underline{0.446} & 0.278 & 0.667 & 0.785 & 0.356 & 0.490 & 0.490 \\ 
LexMAE & 110M & \underline{0.763} & 0.347 & 0.710 & 0.159 & - & 0.500 & 0.219 & 0.562 & - & {0.424} & \underline{0.290} & \textbf{0.716} & \underline{0.800} & 0.352 & - & 0.487 \\ 
DRAGON+ & 110M & 0.759 & 0.339 & 0.679 & 0.159 & 0.875 & 0.469 & 0.227 & 0.537 & 0.354 & 0.414 & 0.263 & 0.662 & 0.781 & 0.359 & 0.491 & 0.491 \\ 
SpladeV3 & 110M & 0.748 & \underline{0.357} & 0.710 & 0.158 & 0.814 & 0.509 & 0.233 & \textbf{0.586} & - & \textbf{0.450} & \textbf{0.293} & \underline{0.692} & 0.796 & 0.374 & - & \underline{0.517} \\ 
SpladeV2 & 110M & 0.710 & 0.334 & 0.693 & 0.158 & 0.838 & 0.479 & 0.235 & 0.521 & 0.341 & 0.435 & 0.272 & 0.684 & 0.786 & 0.336 & 0.487 & 0.487 \\ 
RetroMae & 110M & \textbf{0.772} & 0.308 & 0.653 & 0.133 & 0.847 & 0.433 & 0.232 & 0.518 & 0.297 & 0.356 & 0.219 & 0.635 & 0.774 & 0.325 & 0.464 & 0.464 \\ 
GenQ & 220M & 0.610 & 0.310 & 0.644 & 0.143 & 0.830 & 0.493 & 0.175 & 0.358 & 0.347 & 0.328 & 0.182 & 0.534 & 0.669 & 0.308 & 0.424 & 0.424 \\ 
GTR$_\text{Base}$ & 110M & 0.539 & 0.308 & 0.600 & 0.149 & 0.881 & 0.511 & 0.241 & 0.495 & 0.357 & 0.347 & 0.205 & 0.535 & 0.660 & 0.349 & 0.441 & 0.441 \\ 
GTR-Large & 335M & 0.557 & 0.329 & 0.639 & 0.158 & \underline{0.890} & 0.525 & 0.262 & 0.547 & 0.384 & 0.391 & 0.219 & 0.579 & 0.712 & 0.424 & 0.473 & 0.473 \\ 
GTRxl & 1.2B & 0.580 & 0.343 & 0.635 & 0.159 & \underline{0.890} & 0.531 & \underline{0.270} & 0.559 & 0.388 & 0.396 & 0.230 & 0.591 & 0.717 & \underline{0.444} & 0.481 & 0.481 \\ 
GTRxxl & 4.8B & 0.500 & 0.342 & 0.662 & 0.161 & \textbf{0.892} & 0.540 & 0.267 & \underline{0.568} & \underline{0.399} & 0.408 & 0.256 & 0.599 & 0.740 & \textbf{0.467} & 0.486 & 0.486 \\ 
\midrule
{\bf BiCA$_\text{small}$} & 33M & 0.661 & 0.347 & 0.727 & \underline{0.214} & 0.880 & \underline{0.555} & 0.264 & 0.502 & \underline{0.399} & 0.391 & 0.222 & 0.637 & \textbf{0.815} & 0.393 & \underline{0.501} & 0.501 \\ 
{\bf BiCA$_\text{Base}$} & 110M & 0.684 & \textbf{0.378} & \textbf{0.762} & \textbf{0.231} & 0.882 & \textbf{0.571} & \textbf{0.279} & 0.529 & \textbf{0.428} & 0.411 & 0.220 & 0.657 & \textbf{0.815} & 0.407 & \textbf{0.518} & \textbf{0.518} \\ 
\bottomrule
\end{tabular}
\caption{Evaluation on all 14 BEIR tasks in a zero-shot setting using nDCG@10. \textbf{Bold} and \underline{underline} denote the best and second-best scores, respectively.}
\label{combined_results_rotated_headers}
\end{table*}

\section{Methodology}\label{meth}

First, we construct a rich, 2-hop citation neighborhood around a set of seed documents. Second, we perform a novel hard-negative mining technique by converting these citation graphs into dense semantic graphs and performing a series of diverse, stochastic traversals to find documents that are semantically close but not directly relevant. We provide an overview of our entire pipeline in Figure \ref{fig:cite}.

\subsection{Data Curation: 2-Hop Citation Neighborhood Construction}

The foundation of our dataset is a large-scale, localized citation graph. The process begins with a seed collection of PubMed abstracts from the \texttt{uiyunkim-hub/pubmed-abstract} dataset on Hugging Face. Our goal was to generate a final dataset of approximately 20,000 query-positive pairs, each with a corresponding set of high-quality hard negatives. To ensure that our selected corpus of 20,000 documents is a fair representation of the much larger PubMed database we plot the embedding distributions in Appendix~\ref{data-select}, Figure \ref{fig:ed}.

To create a candidate pool for these negatives, we performed the following steps for each seed article, which we designate as the ``positive" document ($P_0$):
\begin{itemize}
    \item \textbf{1-Hop Citation Retrieval:} Using the PubMed Identifier (PMID) of $P_0$, we employed the \texttt{pubmed-parser} library to query the NCBI E-utilities API and retrieve a list of all PMIDs cited by $P_0$. These form the 1-hop neighborhood ($C_1$). We then fetched the abstract for each paper in $C_1$.
    \item \textbf{2-Hop Citation Retrieval:} For each paper $P_1 \in C_1$, we repeated the process, fetching the PMIDs of all papers it cites. This collection of PMIDs forms the 2-hop neighborhood ($C_2$). We then fetched the abstract for each paper in $C_2$.
    \item \textbf{Data Aggregation:} The final curated data structure for each positive document $P_0$ consists of its own abstract, a list of all 1-hop abstracts, and a list of all 2-hop abstracts. To ensure a sufficiently rich neighborhood for mining, we only retained records where abstracts could be successfully retrieved.
\end{itemize}
This data collection was heavily parallelized across 80 worker processes to manage the high volume of API calls to NCBI. The result is a JSONL file containing 20,000 complex objects, each representing a positive document and its extensive 2-hop citation context.

\subsection{Hard-Negative Mining via Diverse Semantic Traversal}

With the 2-hop citation neighborhoods established, we proceeded to the core of our hard-negative mining strategy. To enhance diversity and prevent the model from overfitting to a single type of negative, our approach, detailed in Algorithm~\ref{alg:semantic_traversal_compact}, transforms the structural citation graph into a semantic space and explores it using multiple, stochastic paths.

\begin{algorithm}[H]
\caption{Hard Negative Mining via Diverse Semantic Traversal}
\label{alg:semantic_traversal_compact}
\small 
\begin{algorithmic}[1]
\Require
    \Statex $A_{\text{pos}}$: Abstract of the positive document.
    \Statex $\mathcal{A}_{\text{cands}}$: Set of candidate abstracts from citation hops.
    \Statex $N_{\text{paths}}, L_{\text{path}}, K_{\text{sample}}$: Traversal control parameters.
\Ensure
    \Statex $L_{\text{negs}}$: A diverse list of hard negative abstracts.
\Procedure{MineHardNegatives}{$A_{\text{pos}}, \mathcal{A}_{\text{cands}}$}
    \State \Comment{Step 1: Construct a semantic graph of documents.}
    \State $Q \gets \Call{GenerateQuery}{A_{\text{pos}}}$
    \State $S_{\text{graph}} \gets \Call{BuildSimilarityGraph}{\mathcal{A}_{\text{cands}}}$
    
    \State \Comment{Step 2: Initiate N traversals from query-relevant starts.}
    \State $I_{\text{start}} \gets \Call{FindTopNStarts}{Q, \mathcal{A}_{\text{cands}}, N_{\text{paths}}}$
    \State $L_{\text{negs}} \gets \emptyset$, $V_{\text{visited}} \gets \emptyset$
    \State \Comment{Step 3: Perform stochastic walks to find diverse negatives.}
    \For{each $i_{\text{start}}$ in $I_{\text{start}}$}
        \State $i_{\text{curr}} \gets i_{\text{start}}$
        \For{$l \gets 1$ to $L_{\text{path}}$}
            \If{$i_{\text{curr}} \in V_{\text{visited}}$} \textbf{break} \EndIf
            \State Add $\mathcal{A}_{\text{cands}}[i_{\text{curr}}]$ to $L_{\text{negs}}$ and $V_{\text{visited}}$
            
            \State \Comment{Select next node: top-K unvisited neighbors,}
            \Statex \hspace*{1.8cm} \Comment{sampled probabilistically by similarity.}
            
            \State $I_{\text{topK}} \gets \Call{GetTopKUnvisitedNeighbors}{}$
            \Statex \hspace*{3.5cm} $\{i_{\text{curr}}, S_{\text{graph}}, K_{\text{sample}}, V_{\text{visited}}\}$
            
            \State $i_{\text{curr}} \gets \Call{SampleProbabilistically}{}$
            \Statex \hspace*{3.5cm} $\{I_{\text{topK}}, S_{\text{graph}}[i_{\text{curr}}, I_{\text{topK}}]\}$
        \EndFor
    \EndFor
    
    \State \Comment{Step 4: Add a random negative for robustness.}
    \State Add one random, unvisited abstract to $L_{\text{negs}}$.
    \State \Return $\operatorname{Unique}(L_{\text{negs}})$
\EndProcedure
\end{algorithmic}
\end{algorithm}

The mining process unfolds in three steps for each of the 20,000 curated data points:
\begin{itemize}
    \item \textbf{Query Generation:} We first generate a synthetic query from the positive abstract ($A_{\text{positive}}$) using the Doc2Query (\texttt{doc2query/all-t5-base-v1}) model~\cite{nogueira_document_2019}. This creates a realistic search query that the positive document is expected to be relevant for.

    \item \textbf{Dense Graph Construction:} We then construct a dense, semantically-weighted graph. All abstracts from the 1-hop and 2-hop neighborhoods are encoded into high-dimensional vectors using the \texttt{Pubmedbert-base-embeddings} \cite{NeuMLpub}. We compute a complete pairwise cosine similarity matrix between all documents in the 1-hop and 2-hop pools.

    \item \textbf{Diverse Semantic Traversal:} With the dense graph constructed, we identify a varied set of hard negatives. The process is designed to be robust and avoid overfitting:
    \begin{itemize}
        \item \textbf{Multiple Start Points:} Instead of one, we initiate three separate traversal paths, starting from the three 1-hop documents most semantically similar to the generated query.
        \item \textbf{Stochastic Path Selection:} At each step of a traversal, rather than taking a purely greedy step to the single most similar node, we perform weighted random sampling from the top five most similar, unvisited nodes. This stochasticity ensures a wider exploration of the semantic space.
        \item \textbf{Global Visited Set:} A single global set of visited nodes is maintained across all traversals for a given query, guaranteeing that each path explores unique documents and maximizing the diversity of the final negative set.
        \item \textbf{Random Negative Augmentation:} Finally, to further improve training stability, one additional negative is selected uniformly at random from the remaining pool of unvisited documents.
    \end{itemize}
\end{itemize}

The final output is a dataset of approximately 20,000 entries, each containing a query, a single positive abstract, and a diverse list of hard negatives (averaging 6.5 per query). This results in a total corpus of approximately 150,000 documents, specifically curated to train and evaluate retrieval models on their ability to make fine-grained relevance distinctions.

\section{Experiments}

\subsection{Fine-Tuning}

We fine tune two models the GTE$_\text{small}$ and the GTE$_\text{Base}$ \cite{li2023generaltextembeddingsmultistage}. GTE\textsubscript{base} (110M parameters, 768‑dim) and GTE\textsubscript{small} (33M parameters, 384‑dim) are BERT‑based embedding models trained with multi‑stage contrastive learning, balancing accuracy with efficiency. We describe our choice of fine-tuning for only 20 steps in Section~\ref{fts} show in Figure \ref{fig:steps} of the Appendix.

The fine-tuning was conducted on a single NVIDIA V100 GPU (32GB), enabling efficient handling of large batch sizes and complex models without memory constraints. The Multiple Negative Ranking Loss (MNR) function \cite{henderson2017efficient} is used and defined as:
{\small
\[
\mathcal{L}_{MNR} = 
-\log \left( 
\frac{\exp(\mathbf{q} \cdot \mathbf{d}_+)}
{\exp(\mathbf{q} \cdot \mathbf{d}_+) + \sum_{i=1}^{K} \exp(\mathbf{q} \cdot \mathbf{d}_i^-)}
\right)
\]
}
where $\mathbf{q}$ denotes the query embedding, $\mathbf{d}_+$ the positive document embedding, $\mathbf{d}_i^-$ the $i$-th negative document embedding, and $K$ the number of negatives.

\section{Evaluation}

\subsubsection{BEIR}

We evaluate our models on fourteen BEIR \cite{thakur_beir_2021} datasets in a zero-shot setting. Details of the dataset is provided in Appendix~\ref{deta}, Table~\ref{tab:beir-datasets}. Our primary evaluation metric is Normalized Discounted Cumulative Gain at 10 (nDCG@10), which assesses the ranking quality of the top 10 retrieved documents. The comprehensive results, comparing our models against a wide range of existing methods, are presented in Table~\ref{combined_results_rotated_headers}. We also provide the improvements over the base GTE models in Appendix \ref{improve} and in Appendix Table \ref{tab:gte_vs_bica}.

As shown in Table~\ref{combined_results_rotated_headers}, our BiCA$_\text{Base}$ model (110M parameters) achieves the highest average nDCG@10 score of 0.518 across all fourteen tasks, setting a new state-of-the-art on BEIR and surpassing significantly larger models such as GTR\_xxl (4.8B parameters, 0.486). BiCA$_\text{Base}$ excels in both biomedical and general domains, leading on \textsc{Nfcorpus} (0.378), \textsc{Scifact} (0.762), \textsc{Scidocs} (0.231), \textsc{ArguAna} (0.571), \textsc{Climate-Fever} (0.279), and \textsc{CQADup} (0.428), while tying for the highest on \textsc{FEVER} (0.815) and performing strongly on \textsc{HotpotQA} (0.657). Our smaller BiCA$_\text{small}$ model (33M parameters) also demonstrates remarkable performance, achieving an average nDCG@10 of {0.501}, ranking second overall and outperforming many larger baselines, including GTR\_xxl. Notably, it secures the top score on \textsc{FEVER} (0.815) and second-highest on \textsc{Scidocs} ({0.214}), \textsc{ArguAna} ({0.555}), and \textsc{CQADup} ({0.399}). Its ability to rival or surpass models up to 145 times larger highlights the parameter efficiency of our approach.

\begin{table*}[ht]
  \centering
  \small
  \begin{tabular}{lcccccc|cc}
    \toprule
    Corpus & ColBERT & BM25 & ANCE & RocketQAv2 & SPLADEv2 & ColBERTv2 & BiCA$_\text{small}$ & BiCA$_\text{Base}$\\
    \midrule
    \multicolumn{7}{l}{\textbf{LoTTE Search Test Queries (Success@5)}}\\
    \midrule
    Writing    & 74.7 & 60.3 & 74.4 & 78.0 & 77.1 & \underline{80.1} & 79.8 & \textbf{81.6}\\
    Recreation & 68.5 & 56.5 & 64.7 & 72.1 & 69.0 & 72.3 & \underline{76.1} & \textbf{79.7}\\
    Science    & 53.6 & 32.7 & 53.6 & 55.3 & 55.4 & 56.7 & \underline{58.5} & \textbf{60.6} \\
    Lifestyle  & 80.2 & 63.8 & 82.3 & 82.1 & 82.3 & 84.7 & \underline{86.8} & \textbf{87.7} \\
    \midrule
    \multicolumn{7}{l}{\textbf{LoTTE Forum Test Queries (Success@5)}}\\
    \midrule
    Writing    & 71.0 & 64.0 & 68.8 & 71.5 & 73.0 & {76.3} & \underline{78.1} & \textbf{80.8}\\
    Recreation & 65.6 & 55.4 & 63.8 & 65.7 & 67.1 & 70.8 & \underline{75.6}& \textbf{77.5} \\
    Science    & 41.8 & 37.1 & 36.5 & 38.0 & 43.7 & \underline{46.1} & 44.6 & \textbf{47.1} \\
    Lifestyle  & 73.0 & 60.6 & 73.1 & 73.7 & 74.0 & 76.9 & \underline{82.2} & \textbf{84.0}\\
    \bottomrule
  \end{tabular}
  \caption{Retrieval performance (Success@5) of different models on LoTTE search and forum queries on the test set. \textbf{Bold} represents the best score and \underline{underline} represents the second best score.}
  \label{tab:lotte-results}
\end{table*}

\begin{table*}[!ht]
\centering
\begin{tabular}{lccccccl}
\toprule
\multirow{2}{*}{\textbf{Model}} & \multirow{2}{*}{\textbf{Batch Size}} & \multicolumn{2}{c}{\textbf{Encoding (ms)}$\downarrow$} & \multicolumn{2}{c}{\textbf{Retrieval (ms)}$\downarrow$} & \multicolumn{2}{c}{\textbf{Total (ms)}$\downarrow$} \\
\cmidrule(lr){3-4} \cmidrule(lr){5-6} \cmidrule(lr){7-8}
& & \textbf{Avg.} & \textbf{99th p.} & \textbf{Avg.} & \textbf{99th p.} & \textbf{Avg.} & \textbf{99th p.} \\
\midrule
\multirow{3}{*}{BiCA$_\text{Base}$} 
& 1    & \cellcolor{myorange2}9   & \cellcolor{myorange3}14  & \cellcolor{myorange2}7   & \cellcolor{myorange4}9   & \cellcolor{myorange3}16  & \cellcolor{myorange5}21 \\
& 10   & \cellcolor{myorange1}11  & \cellcolor{myorange3}16  & \cellcolor{myorange2}9   & \cellcolor{myorange2}10  & \cellcolor{myorange2}20  & \cellcolor{myorange3}25 \\
& 2000 & \cellcolor{myorange5}1292& \cellcolor{myorange5}1475& \cellcolor{myorange5}612 & \cellcolor{myorange5}622 & \cellcolor{myorange5}1904& \cellcolor{myorange5}2082\\
\midrule
\multirow{3}{*}{BiCA$_\text{small}$} 
& 1    & \cellcolor{myorange2}9   & \cellcolor{myorange2}11  & \cellcolor{myorange1}4   & \cellcolor{myorange1}4   & \cellcolor{myorange1}13  & \cellcolor{myorange1}14 \\
& 10   & \cellcolor{myorange2}14  & \cellcolor{myorange4}19  & \cellcolor{myorange1}5   & \cellcolor{myorange1}5   & \cellcolor{myorange1}19  & \cellcolor{myorange2}24 \\
& 2000 & \cellcolor{myorange1}554 & \cellcolor{myorange1}850 & \cellcolor{myorange1}441 & \cellcolor{myorange1}504 & \cellcolor{myorange1}994 & \cellcolor{myorange1}1341\\
\midrule
\multirow{3}{*}{ColBERTv2} 
& 1    & \cellcolor{myorange1}8   & \cellcolor{myorange1}9   & \cellcolor{myorange2}7   & \cellcolor{myorange2}7   & \cellcolor{myorange2}15  & \cellcolor{myorange2}16 \\
& 10   & \cellcolor{myorange1}11  & \cellcolor{myorange1}13  & \cellcolor{myorange2}9   & \cellcolor{myorange2}10  & \cellcolor{myorange2}20  & \cellcolor{myorange1}23 \\
& 2000 & \cellcolor{myorange3}1249& \cellcolor{myorange3}1423& \cellcolor{myorange3}594 & \cellcolor{myorange4}612 & \cellcolor{myorange3}1844& \cellcolor{myorange3}2004\\
\midrule
\multirow{3}{*}{RetroMAE} 
& 1    & \cellcolor{myorange2}9   & \cellcolor{myorange2}11  & \cellcolor{myorange2}7   & \cellcolor{myorange3}8   & \cellcolor{myorange3}16  & \cellcolor{myorange4}20 \\
& 10   & \cellcolor{myorange1}11  & \cellcolor{myorange1}13  & \cellcolor{myorange2}9   & \cellcolor{myorange3}12  & \cellcolor{myorange2}20  & \cellcolor{myorange3}25 \\
& 2000 & \cellcolor{myorange2}1246& \cellcolor{myorange2}1403& \cellcolor{myorange2}591 & \cellcolor{myorange2}607 & \cellcolor{myorange2}1837& \cellcolor{myorange2}1985\\
\midrule
\multirow{3}{*}{SpladeV3} 
& 1    & \cellcolor{myorange2}9   & \cellcolor{myorange2}11  & \cellcolor{myorange2}7   & \cellcolor{myorange4}9   & \cellcolor{myorange3}16  & \cellcolor{myorange3}19 \\
& 10   & \cellcolor{myorange1}11  & \cellcolor{myorange2}15  & \cellcolor{myorange2}9   & \cellcolor{myorange4}13  & \cellcolor{myorange3}21  & \cellcolor{myorange4}32 \\
& 2000 & \cellcolor{myorange4}1250& \cellcolor{myorange4}1437& \cellcolor{myorange4}598 & \cellcolor{myorange3}609 & \cellcolor{myorange4}1847& \cellcolor{myorange4}2045\\
\bottomrule
\end{tabular}
\caption{Latency analysis for BiCA$_\text{Base}$, BiCA$_\text{small}$, and other baselines on a V100 (32GB) GPU. Cell colors highlight timings from lowest (lightest orange) to highest (darkest orange) for each metric across models within the same batch size. All times are in milliseconds (ms). \textit{Encoding} refers to query encoding time, and \textit{Retrieval} to top-1000 passage retrieval from a FAISS index with 10,000 passages (MS MARCO).}
\label{tab:latency_results_colored_rowwise_orange}
\end{table*}

\subsubsection{LOTTE}

We evaluate our models on long-tailed topics, which refer to specific and less frequently searched queries, using four sub-topics from the LoTTE benchmark~\cite{santhanam-etal-2022-colbertv2}: \textit{Science}, \textit{Writing}, \textit{Recreation}, and \textit{Lifestyle}.Details of the dataset is provided in Appendix~\ref{deta}, Table~\ref{tab:lotte-composition}. As detailed in Table~\ref{tab:lotte-results}, we report zero-shot Success@5 on its test set. The benchmark includes two query formats: concise Search queries from GooAQ logs and more descriptive Forum queries from StackExchange user questions.

Our BiCA$_\text{Base}$ model sets a new state-of-the-art, achieving the highest Success@5 across all four categories for both LoTTE query types. On Search queries, it scores 87.7 on \textit{Lifestyle}, 81.6 on \textit{Writing}, 79.7 on \textit{Recreation}, and 60.6 on \textit{Science}. On the more challenging Forum queries, it attains 84.0 on \textit{Lifestyle}, 80.8 on \textit{Writing}, 77.5 on \textit{Recreation}, and 47.1  on \textit{Science}. The smaller BiCA$_\text{small}$ model consistently ranks second, with Search scores of 86.8 on \textit{Lifestyle}, 76.1 on \textit{Recreation} and 58.5 on \textit{Science}, and Forum scores of 82.2 on \textit{Lifestyle}, 78.1 on \textit{Writing} and 75.6 on \textit{Recreation}, demonstrating strong performance and parameter efficiency on long-tailed topics.

\subsubsection{Latency}
\label{sec:latency}

To assess model efficiency, we measured latency using the TAS-B setup on a single NVIDIA V100 with 32GB memory. We encoded 10,000 MS MARCO passages and indexed them with FAISS (IndexFlatIP). We then timed two steps: query encoding and retrieval of top 1000 results. Tests were run on query batches of size 1, 10, and 2000. We report average and 99th percentile latencies in milliseconds over 100 iterations (1 and 10) or 10 iterations (2000).

Table~\ref{tab:latency_results_colored_rowwise_orange} compares BiCA$_\text{Base}$ (110M), BiCA$_\text{small}$ (33M), ColBERTv2, RetroMAE, and SpladeV3. For batch size 1, BiCA$_\text{small}$ is fastest overall with 13\,ms total and 4\,ms retrieval. ColBERTv2 has the quickest encoding at 8\,ms and a total of 15\,ms. The others average 16\,ms, with BiCA$_\text{Base}$ showing slightly higher tail times.

At batch size 10, BiCA$_\text{small}$ again leads in total time (19\,ms), driven by retrieval at 5\,ms. ColBERTv2, RetroMAE, and SpladeV3 encode slightly faster (11\,ms vs 14\,ms for BiCA$_\text{small}$). ColBERTv2 has the best tail latency at 23\,ms, while SpladeV3 peaks at 32\,ms.

At batch size 2000, BiCA$_\text{small}$ outperforms all others with 994\,ms total (554\,ms encoding, 441\,ms retrieval). RetroMAE follows at 1837\,ms, then ColBERTv2 (1844\,ms) and SpladeV3 (1847\,ms). BiCA$_\text{Base}$ is slowest at 1904\,ms.

\section{Effect of Traversal Parameters}

To determine appropriate values for the traversal parameters, we conduct an ablation study varying the \textit{Number of Traversal Paths} ($N_{paths}$) and the \textit{Length of the Path} ($L_{path}$) in the range of 1–5. For each study, we fix one parameter at 3 while varying the other, using a \texttt{bert-base} fine-tuned for 1 epoch on the entire corpus with a batch size of 16 and an MNR loss. As shown in Table~\ref{tab:main_results}, the choice of $N_{paths}=3$ and $L_{path}=3$ consistently provides a strong balance across datasets, achieving the highest overall average performance (0.2739). While other configurations occasionally yield the best score on a single dataset (e.g., $N_{paths}=5$ for SCIFACT or $L_{path}=1$ for ArguAna), they underperform on others, leading to a lower overall average. We therefore select $N_{paths}=3$ and $L_{path}=3$ as the default configuration for our final results, as it offers the most stable and robust performance across benchmarks.

\begin{table*}[t!]
\centering
\begin{tabular}{ccccccc|c}
\toprule
\textbf{$N_{paths}$} & \textbf{$L_{path}$} & \textbf{NFC} & \textbf{SCIDOCS} & \textbf{SCIFACT} & \textbf{ArguAna} & \textbf{FIQA} & \textbf{Average} \\
\midrule
\multicolumn{8}{l}{\textit{Ablation on Number of Traversals (fixed $L_{path}=3$)}} \\
\midrule
1 & 3 & 0.1803 & 0.1201 & 0.5114 & 0.3974 & 0.1301 & 0.2679 \\
2 & 3 & 0.1390 & 0.0984 & 0.3934 & 0.3174 & 0.0860 & 0.2068 \\
4 & 3 & 0.1400 & 0.1073 & 0.4392 & 0.3024 & 0.1030 & 0.2184 \\
5 & 3 & 0.1891 & 0.1230 & \textbf{0.5180} & 0.4190 & 0.1178 & 0.2734 \\
\midrule
\multicolumn{8}{l}{\textit{Ablation on Path Length (fixed $N_{paths}=3$)}} \\
\midrule
3 & 1 & 0.1875 & \textbf{0.1245} & 0.5053 & \textbf{0.4211} & 0.1240 & 0.2725 \\
3 & 2 & 0.1299 & 0.0965 & 0.3960 & 0.2920 & 0.1062 & 0.2041 \\
\rowcolor{yellow}
3 & 3 & \textbf{0.1987} & 0.1234 & 0.5156 & 0.4094 & 0.1225 & \textbf{0.2739} \\
3 & 4 & 0.1861 & 0.1202 & 0.5102 & 0.3854 & \textbf{0.1324} & 0.2669 \\
3 & 5 & 0.1820 & 0.1183 & 0.5058 & 0.3730 & 0.1110 & 0.2580 \\
\bottomrule
\end{tabular}%
\caption{Ablation study on the number of traversals ($N_{paths}$) and path length ($L_{path}$). All models are based on BERT-base fine-tuned for one epoch. We report NDCG@10 scores and highlight the best result in each column in \textbf{bold}.}
\label{tab:main_results}
\end{table*}

\begin{table}[h!]
\centering
\small
\setlength{\tabcolsep}{4pt} 
\renewcommand{\arraystretch}{0.95} 
\begin{tabular}{l@{\hskip 6pt}c@{\hskip 6pt}c@{\hskip 6pt}c@{\hskip 6pt}c@{\hskip 6pt}c@{\hskip 6pt}c}
\toprule
\textbf{Dataset} & \textbf{Baseline} & \textbf{1k} & \textbf{5k} & \textbf{10k} & \textbf{15k} & \textbf{Full (20k)} \\
\midrule
NFCorpus         & 0.043             & 0.082       & 0.171       & 0.164        & 0.171        & \textbf{0.185}      \\
SciDocs          & 0.028             & 0.061       & 0.117       & 0.116        & 0.114        & \textbf{0.121}      \\
SciFact          & 0.130             & 0.262       & 0.469       & 0.468        & 0.492        & \textbf{0.493}      \\
ArguAna          & 0.283             & 0.384       & 0.364       & 0.385        & 0.405        & \textbf{0.444}      \\
\bottomrule
\end{tabular}
\caption{Scaling ablation results for fine-tuning \texttt{bert-base-uncased} on our citation-aware negatives. Scores are nDCG@10 on biomedical BEIR tasks. The baseline represents zero-shot performance without any fine-tuning. The results show consistent performance improvement as the amount of training data increases.}
\label{tab:bert-scaling}
\end{table}

\begin{table}[!ht]
\small
\setlength{\tabcolsep}{6pt} 
\centering
\begin{tabularx}{0.95\linewidth}{l|cc|cc}
\toprule
\textbf{Dataset} & 
\textbf{{DB\textsubscript{base}}} & 
\textbf{{DB\textsubscript{fine-tune}}} & 
\textbf{{E5\textsubscript{base}}} & 
\textbf{{E5\textsubscript{fine-tune}}} \\
\midrule
NFCorpus   & 24.8 & 25.2\textsuperscript{+0.4} & 35.3 & 34.8\textsuperscript{--0.5} \\
SciFact    & 51.6 & 55.8\textsuperscript{+4.2} & 71.0 & 71.9\textsuperscript{+0.9} \\
SCIDOCS    & 13.4 & 14.9\textsuperscript{+1.5} & 18.3 & 20.4\textsuperscript{+2.1} \\
ArguAna    & 39.7 & 39.9\textsuperscript{+0.2} & 51.6 & 52.7\textsuperscript{+1.1} \\
FiQA       & 18.2 & 19.7\textsuperscript{+1.5} & 37.3 & 37.9\textsuperscript{+0.6} \\
\midrule
\textbf{Average $\Delta$} & -- & \textbf{+1.56} & -- & \textbf{+0.84} \\
\bottomrule
\end{tabularx}
\caption{NDCG@10 (\%) comparison between \texttt{DistilBERT (DB)} and \texttt{E5} models across BEIR datasets. Superscripts indicate absolute improvement of fine-tuned models over base versions.}
\label{DA}
\end{table}

\begin{table}[!ht]
    \centering
    \begin{tabular}{cc}
    \toprule
    \textbf{Model} & \textbf{No. Fine-Tuning Steps} \\
    \midrule
    {DistilBERT} & 1150 \\
    {e5-base-v2} & 290 \\
    \bottomrule
    \end{tabular}
    \caption{Number fine-tuning steps on our constructed corpus before doing zero-shot evaluation on BEIR}
    \label{tab:MFTS}
\end{table}

\section{Robustness and Scalability}
\label{sec:robustness}

To examine the effect of training data size, we fine-tuned \texttt{bert-base-uncased} \cite{devlin-etal-2019-bert} on randomly sampled subsets of our 20,000-record dataset (1k, 5k, 10k, 15k, and full). Each subset reserved 10\% for validation. Models were trained for up to 1 epoch using MNR Loss with a batch size of 16, applying early stopping based on highest triplet accuracy on validation. The best checkpoints were evaluated zero-shot on three biomedical tasks and one BEIR task. Results in Table~\ref{tab:bert-scaling} show a clear positive correlation between data size and retrieval performance.

\section{Performance of Different Architectures}

To assess generalizability, we fine-tune models for a maximum of one epoch with early stopping (patience=3), where evaluation is performed every 10 steps. We experiment with two pretrained checkpoints: e5-base-V2\footnote{\url{https://huggingface.co/intfloat/e5-base-v2}} \cite{wang2022text} and a DistilBERT model\footnote{\url{https://huggingface.co/GPL/msmarco-distilbert-margin-mse}} \cite{sanh_distilbert_2020} fine-tuned on MS MARCO. For evaluation, we select five tasks from the BEIR benchmark three from the biomedical domain (NFCorpus, SciDocs, SciFact) and two from non-biomedical domains (ArguAna, FiQA). Table \ref{DA} shows the performance gains of fine-tuning the models on our corpus and Table \ref{tab:MFTS} shows the number of fine tuning steps selected for the chosen models, after which we do zero-shot evaluation on BEIR.

We see consistent improvements in using our corpus for fine-tuning over different architectures. DistilBERT sees an average improvement of 1.56 points and e5-base-v2 sees an improvement of 0.84 points.

\section{Conclusions}

In this work, we present BiCA$_\text{Base}$ and BiCA$_\text{small}$, two dense retrieval models designed to address the unique challenges of biomedical and general-domain information retrieval. At the core of our approach is a novel hard negative mining strategy that exploits multi-hop citation chains extracted from PubMed. This citation-aware technique provides semantically challenging yet relevant negative examples, encouraging the models to learn fine-grained distinctions essential for high-precision retrieval.

Through extensive experiments on the BEIR benchmark, BiCA$_\text{Base}$ demonstrated strong performance across both biomedical and non-biomedical tasks, consistently outperforming several larger state-of-the-art models. Notably, it achieved the highest average nDCG@10 scores in both domains, indicating its effectiveness and generalizability. Despite its smaller size, BiCA$_\text{small}$ also delivered competitive results, often closely trailing BiCA$_\text{Base}$ while offering substantially lower inference latency, making it well-suited for real-time and resource-constrained applications.

Evaluations on the LoTTE dataset further highlighted the robustness of our models in handling retrieval over long-tailed, diverse topics. BiCA$_\text{Base}$ led across all sub-domains, while BiCA$_\text{small}$ ranked consistently among the top performers, demonstrating the broad applicability and efficiency of our approach.

\section{Limitations}
The citation-aware hard negative mining strategy improves retrieval performance, but faces challenges in scalability and efficiency. Constructing multi-hop citation chains requires iterative PubMed API requests for abstracts and cited PMIDs, a process hindered by rate limits, network latency, and the parsing of large text data. As a result, generating large training sets can take week(s), depending on the number of seed documents and citation depth. Furthermore, our current work is restricted to PubMed; extending this approach to other sources such as Wikipedia, where scientific and technical articles contain rich citation trails, may enable construction of semantically meaningful hard negatives for general-domain retrieval while preserving citation-aware principles. We acknowledge that our latency evaluation setup may not fully reflect the efficiency advantages of the ColBERTv2 model. However, we adopted this configuration to ensure a uniform and fair comparison across all systems.
\bibliography{main}
\newpage
\appendix
\setcounter{secnumdepth}{2}
\section{Improvement over GTE Models}
\label{improve}

Table~\ref{tab:gte_vs_bica} presents the retrieval performance comparison between our BiCA models and the corresponding GTE \cite{li2023generaltextembeddingsmultistage} baselines across fourteen datasets. BiCA$_\text{small}$ achieves consistent improvements over GTE$_\text{small}$, with an average gain of $\sim$5.8 points. BiCA$_\text{Base}$ shows even stronger results, outperforming GTE$_\text{Base}$ by an average of $\sim$6.8 points. These gains highlight the effectiveness of BiCA’s training strategy in enhancing retrieval quality, particularly on challenging datasets such as ArguAna, NQ, HotpotQA, and Climate-Fever.

\begin{table}[!ht]
\centering
\small
\begin{tabular}{lcc|cc}
\toprule
Dataset        & GTE$_\text{small}$ & BiCA$_\text{Small}$ & GTE$_\text{base}$ & BiCA$_\text{Base}$ \\
\midrule
ArguAna        & 41.6 & 55.5$^{+13.9}$ & 41.0 & 57.1$^{+16.1}$ \\
Climate‐Fever  & 21.4 & 26.4$^{+5.0}$  & 21.0 & 27.9$^{+6.9}$ \\
CQADupStack    & 38.1 & 39.9$^{+1.8}$  & 39.9 & 42.8$^{+2.9}$ \\
DBPedia        & 33.5 & 39.1$^{+5.6}$  & 33.2 & 41.1$^{+7.9}$ \\
Fever          & 71.3 & 81.5$^{+10.2}$ & 72.7 & 81.5$^{+8.8}$ \\
FiQA           & 37.0 & 39.3$^{+2.3}$  & 36.9 & 40.7$^{+3.8}$ \\
HotpotQA       & 49.3 & 63.7$^{+14.4}$ & 50.8 & 65.7$^{+14.9}$ \\
NFCorpus       & 34.9 & 34.7$^{-0.2}$  & 36.2 & 37.8$^{+1.6}$ \\
NQ             & 32.0 & 50.2$^{+18.2}$ & 35.3 & 52.9$^{+17.6}$ \\
Quora          & 86.1 & 88.0$^{+1.9}$  & 85.0 & 88.2$^{+3.2}$ \\
Scidocs        & 21.5 & 21.4$^{-0.1}$  & 22.5 & 23.1$^{+0.6}$ \\
SciFact        & 72.7 & 72.7$^{0.0}$   & 74.1 & 76.2$^{+2.1}$ \\
Touché-2020    & 17.7 & 22.2$^{+4.5}$  & 18.2 & 22.0$^{+3.8}$ \\
Trec‐Covid     & 61.8 & 66.1$^{+4.3}$  & 64.0 & 68.4$^{+4.4}$ \\
\midrule
\textbf{Average $\Delta$} & -- & \textbf{+5.8} & -- & \textbf{+6.8} \\
\bottomrule
\end{tabular}
\caption{Comparison of GTE$_\text{small}$/GTE$_\text{base}$ vs.\ BiCA$_\text{Small}$/BiCA$_\text{Base}$ on 14 tasks. All scores have been multiplied by 100, and the gain next to each BiCA score is rounded to one decimal. The last row reports the average gain across tasks.}
\label{tab:gte_vs_bica}
\end{table}

\section{Data Selection}
\label{data-select}

\begin{figure}[!ht]
    \centering
    \begin{subfigure}[t]{\linewidth}
        \centering
        \includegraphics[width=\linewidth]{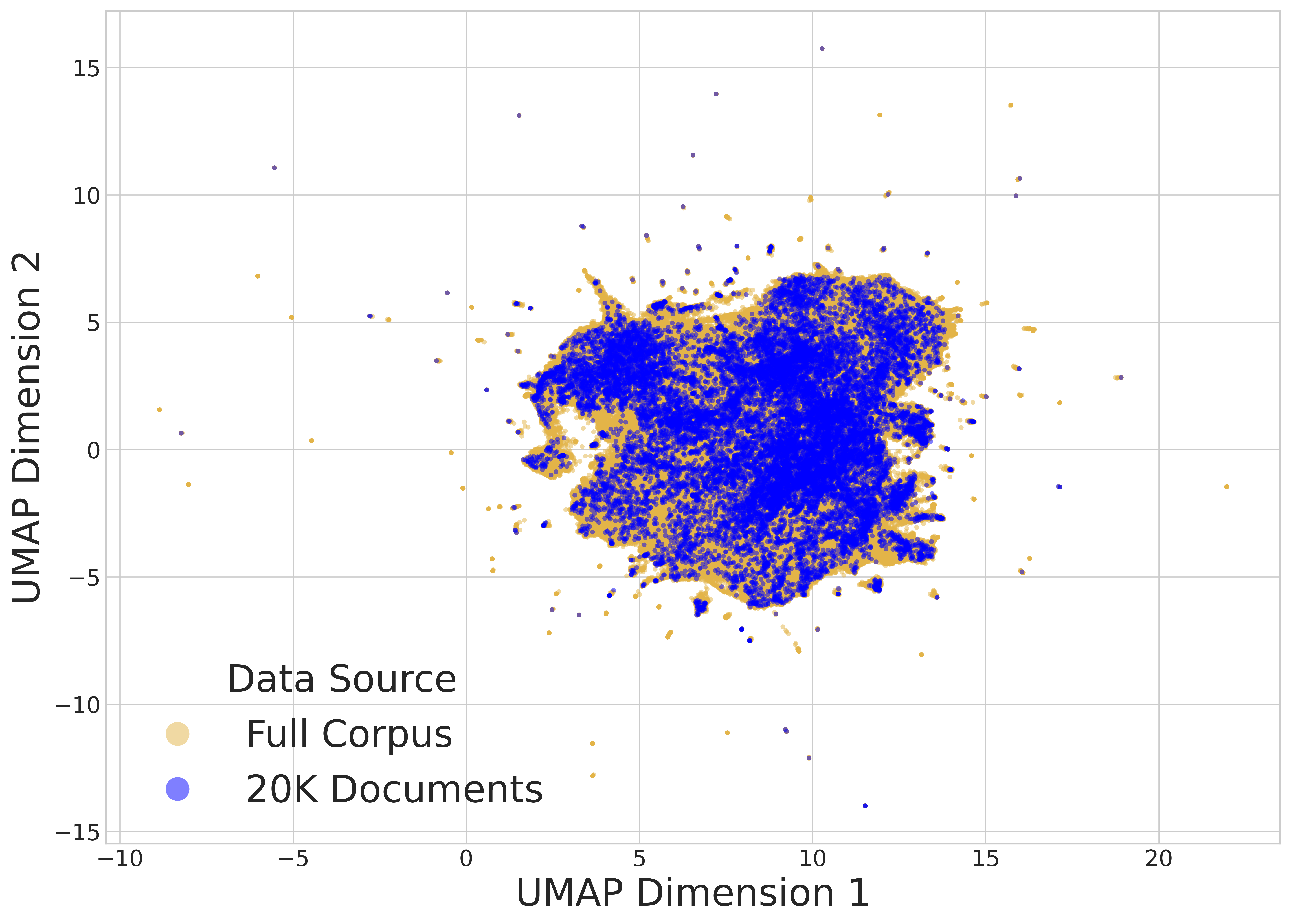}
        \caption{Embedding distributions of the entire corpus (yellow) vs the selected 20,000 documents (blue) to build our training corpus.}
        \label{fig:ed}
    \end{subfigure}
    \\
    \begin{subfigure}[t]{\linewidth}
        \centering
        \includegraphics[width=\linewidth]{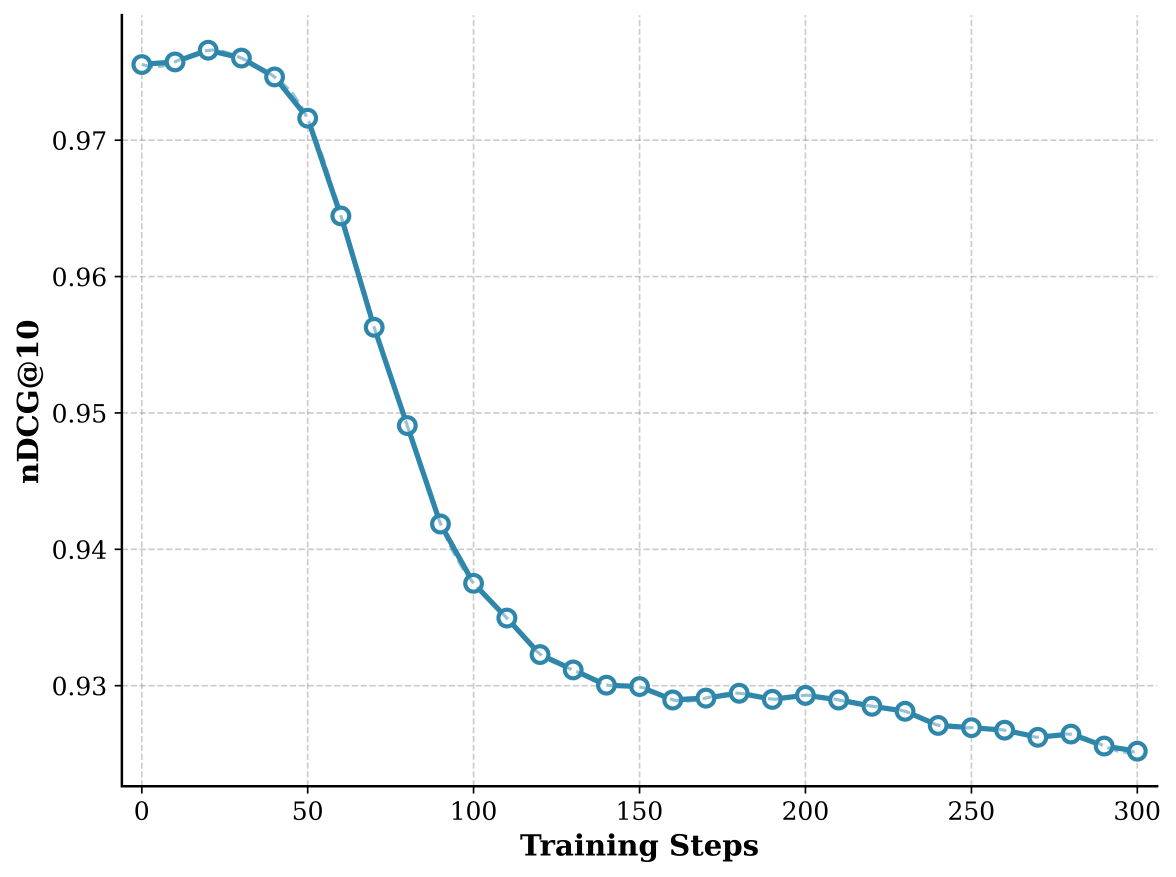}
        \caption{nDCG@10 scores on the validation set using an 80\%/20\% split of the constructed corpus. Evaluation is done every 10 steps, with peak performance observed at step 20 selected for full fine-tuning on the entire corpus followed by zero-shot evaluation on BEIR.}
        \label{fig:steps}
    \end{subfigure}
    \caption{(a) Corpus embedding distribution comparison and (b) validation nDCG@10 across training steps.}
    \label{fig:combined}
\end{figure}

To ensure that our selected training corpus of 20,000 documents is representative of the entire dataset~\footnote{\url{huggingface.co/datasets/uiyunkim-hub/pubmed-abstract}} that was available we plot the distribution of our corpus and the entire corpus as seen in Figure \ref{fig:ed}. We use the \textit{NeuML/pubmedbert-base-embeddings-2M}~\footnote{\url{huggingface.co/NeuML/pubmedbert-base-embeddings-2M}} model to extract the embeddings. 

\begin{table*}[h]
\centering
\small
\begin{tabular}{llrrl}
\toprule
\textbf{Topic} & \textbf{Question Set} & \textbf{\# Questions} & \textbf{\# Passages} & \textbf{Subtopics} \\
\midrule
\multirow{2}{*}{Writing} 
  & Search & 1071 & 200k & English \\
  & Forum  & 2000 & 200k & English \\
\midrule
\multirow{2}{*}{Recreation} 
  & Search & 924 & 167k & Gaming, Anime, Movies \\
  & Forum  & 2002 & 167k & Gaming, Anime, Movies \\
\midrule
\multirow{2}{*}{Science} 
  & Search & 617 & 1.694M & Math, Physics, Biology \\
  & Forum  & 2017 & 1.694M & Math, Physics, Biology \\
\midrule
\multirow{2}{*}{Lifestyle} 
  & Search & 661 & 119k & Cooking, Sports, Travel \\
  & Forum  & 2002 & 119k & Cooking, Sports, Travel \\
\bottomrule
\end{tabular}
\caption{Composition of LoTTE showing test topics, question sets, and a sample of corresponding subtopics. Search Queries are taken from GooAQ, while Forum Queries are taken directly from the StackExchange archive.}
\label{tab:lotte-composition}
\end{table*}

\begin{table*}[ht]
\centering
\small
\begin{tabular}{l l r r}
\toprule
\textbf{Dataset} & \textbf{License} & \textbf{\# Passages} & \textbf{\# Test Queries} \\
\midrule
ArguAna \cite{arguana} & CC BY 4.0 & 8674 & 1406 \\
Touché-2020 \cite{bondarenko2020overview} & CC BY 4.0 & 382545 & 49 \\
NFCorpus \cite{ferro_full-text_2016} & Not reported & 3633 & 323 \\
NQ \cite{nq} & CC BY-SA 3.0 & 2681468 & 3452 \\
DBPedia \cite{dbpedia} & CC BY-SA 3.0 & 4635922 & 400 \\
FEVER \cite{thorne-etal-2018-fever} & CC BY-SA 3.0 &  5416568 & 6666 \\
SCIDOCS \cite{cohan_specter_2020} & GNU General Public License v3.0 & 25657 & 1000 \\
SciFact \cite{wadden_fact_2020} & CC BY-NC 2.0 & 5183 & 300 \\
Quora & Not reported & 522931 & 10000 \\
FiQA \cite{fiqa} & Not reported &  57638 & 648 \\
Climate-Fever \cite{diggelmann2021climatefeverdatasetverificationrealworld} & Not reported &  5416593 & 1535 \\
TREC-COVID \cite{voorhees_trec-covid_2021} & Dataset License Agreement & 171332 & 50 \\
CQADupStack \cite{cqa} & Apache License 2.0 & 457199 & 13145 \\
HotPotQA \cite{yang-etal-2018-hotpotqa} &  CC BY-SA 4.0 & 5233329  & 7405 \\
\bottomrule
\end{tabular}
\caption{BEIR dataset information.}
\label{tab:beir-datasets}
\end{table*}

\section{Choice of fine tuning steps}
\label{fts}

To determine the optimal fine-tuning duration, we evaluated performance on a held-out validation set using a 80\%/20\% split of the constructed corpus. As shown in Figure \ref{fig:steps}, we observed that our highly informative negatives deliver their signal with remarkable speed. Peak performance was consistently achieved at just 20 training steps. This demonstrates the extreme efficiency of our citation-aware negatives. Consequently, we selected this optimal 20-step checkpoint for all zero-shot evaluations on BEIR and LoTTE \cite{santhanam-etal-2022-colbertv2}.

\section{Dataset Details}
\label{deta}
\subsection{BEIR}

Table \ref{tab:beir-datasets} lists the BEIR datasets we used in our evaluation of the BiCA models, including their license information as well as the number of documents and queries present in the dataset. For a more detailed description of the datasets we refer to \cite{thakur_beir_2021}.

\subsection{LoTTE test-set}

Table \ref{tab:lotte-composition} details the sub-topics we evaluated the BiCA models on from the LoTTE test-set. We refer the dataset descriptions exactly as they were in \cite{santhanam-etal-2022-colbertv2}.

\end{document}